\documentclass[sigconf,anonymous=false]{acmart}

\usepackage{natbib}
\usepackage[utf8]{inputenc} 
\usepackage[T1]{fontenc}    
\usepackage{hyperref}       
\usepackage{url}            
\usepackage{booktabs}       
\usepackage{amsfonts}       
\usepackage{amsmath}
\usepackage{nicefrac}       
\usepackage{microtype}      
\usepackage{subfig}
\usepackage{graphicx}
\usepackage{pifont}         
\usepackage{float}
\usepackage{lipsum}
\usepackage{graphics}
\usepackage{stfloats}

\newcommand{\cmark}{\ding{51}}%
\newcommand{\xmark}{\ding{55}}%

\AtBeginDocument{%
  \providecommand\BibTeX{{%
    \normalfont B\kern-0.5em{\scshape i\kern-0.25em b}\kern-0.8em\TeX}}}

\setcopyright{acmcopyright}
\copyrightyear{2021}
\acmYear{2021}
\acmDOI{10.1145/1122445.1122456}

\begin{document}

\title{Deep Pairwise Learning To Rank For Search Autocomplete}

\author{Kai Yuan, Da Kuang}
\affiliation{\institution{Amazon Search}}
\email{{yuankai, dakuang}@amazon.com}

\begin{abstract}
Autocomplete (a.k.a ``Query Auto-Completion'', ``AC'') suggests full queries based on a prefix typed by customer. Autocomplete has been a core feature of commercial search engine. In this paper, we propose a novel context-aware neural-network based pairwise ranker (DeepPLTR) to improve AC ranking, DeepPLTR leverages contextual and behavioral features to rank queries by minimizing a pairwise loss, based on a fully-connected neural network structure. Compared to LambdaMART ranker, DeepPLTR shows +3.90\% MeanReciprocalRank (MRR) lift in offline evaluation, and yielded +0.06\% (p < 0.1) Gross Merchandise Value (GMV) lift in an Amazon's online A/B experiment.
\end{abstract}



\maketitle

\section{Introduction}

Autocomplete is a common feature of most modern commercial search engines. It refers to the task of suggesting full queries after customer typed a prefix of a few characters \cite{cai2016survey}. Study shows that AC feature can significantly reduce the number of characters typed \cite{zhang2015adaqac}. An Autocomplete system generally consists of 2 steps: Matching and Ranking. Matching refers to generating query candidates given a typed prefix. Ranking refers to ranking the candidates. In this paper, we focus on the Ranking step and propose a NN-based online deep pairwise learning to rank (DeepPLTR) ranker. Our work makes the following  contributions:

\begin{itemize}
  \item Propose a novel NN-based ranker to rank queries by minimizing pairwise loss.
  \item Propose a novel method to incorporate the contextual signals into ranking.
\end{itemize}

The paper is organized as follows: Section 2 describes the related work in the literature; Section 3 proposes the DeepPLTR model; Section 4 presents the datasets, evaluation methods and experiments; Section 5 summarizes our work and discusses possible future research directions.

\section{Related Work}

A detailed survey of AC work could be found in \cite{cai2016survey}, which categories the Autocomplete into 2 types: heuristic approaches and learning-based approaches. Heuristic approaches use fixed algorithms to compute a ranking score for each query. Learning-based approaches treat the Autocomplete as a ranking problem, and rely on the research of learning-to-rank(LTR) field \cite{liu2009learning}, it requires a large number of training samples and features, and generally outperforms the heuristic approaches \cite{cai2016survey}. Ranking features can be categorized into 3 types: time-senstive features, personalized features and contextual features. Time-sensitive features are based on query popularity and could change over time. Personalized features, such as location, are typically limited and hard to access. Contextual features rely on the customer's previous activity and are easy to extract from the search logs. Incorporating contextual signal into model is challenging. Bar-Yossef and Kraus \cite{bar2011context} use user's previous queries in their context-aware AC. Shokouhi \cite{shokouhi2013learning} used a combination of context-based textual features and personalized features in their context-aware AC, and showed that the user’s long term search history and location are most effective signals. NN-based approach is also widely studied for AC ranking task. Session-based, personalized, and attention-based models were proposed in \cite{jiang2018neural} Fiorini and Lu \cite{fiorini2018personalized} extracted user history based features and time features as an input to an RNN model. Latest approach for AC ranking was proposed by Manojkumar et al. \cite{kannadasan2019personalized}, this approach was to train query and context embeddings on the reformulations behavior in search sessions, calculate contextual similarity features from the context and query embeddings, and then feed contextual features to a LambdaMART \cite{burges2010ranknet} ranker. This embedding approach is similar to the ``whole query embedding'' proposed in this paper when computing contextual similarity features.

\section{Deep Pairwise Learning To Rank Model}
In this section, we first describe a common labeling strategy for pairwise AC ranking task, then demonstrate the DeepPLTR architecture and ranking loss, then we propose a scalable approach to learn whole query embeddings and apply it to generate contextual features for ranking model.

\subsection{Training Sample Generation}

\begin{figure}
\centering
\includegraphics[width=\linewidth]{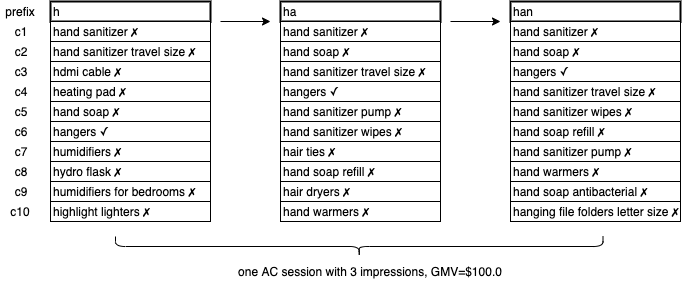}
\caption{Illustration of one AC session: Customer typed keystrokes "h"=>"a"=>"n" and clicked ``hangers'', and then purchased \$100 products.}
\label{fig:Training Generation}
\end{figure}

AC session dataset is the main source to extract training samples. One AC session consists of aggregated impression logs that starts from the first keystroke in search box, all the way until customer submits the query. AC session has been attributed with the right metrics (GMV, sales, clicks etc) to each impression, as the metrics are only observed after customer submits the query, e.g. viewing or purchasing products. We propose a similar labeling strategy: we start by sampling a set of ac sessions. For each ac session, we assume that the query that was eventually submitted is the only right (or the most relevant) query, which should have been suggested right after the first keystroke and all the way until submission. With this assumption in mind, we obtained all the impressed queries from each suggestion; for each pair of <prefix, completion list> constructed this way, we assign positive label to the query submitted by user at the end (if it appears in the list) as {\em relevant keywords} and negative label to other impressed queries as {\em irrelevant keywords}. Then we grouped these positive and negative keywords from the same impression into pairs. Figure \ref{fig:Training Generation} provides an example: here we sampled one series of impression logs in which the user had submitted ``hangers'' as query. For each prefix, the only query that is assigned a positive label is specified with a checkmark (\cmark), and the query that is assigned a negative label is specified with a crossmark (\xmark). The attribution of the pairs is set to 100 as GMV attribution is \$100 for this AC session.

\begin{figure*}[b]
\centering
\includegraphics[scale=0.44]{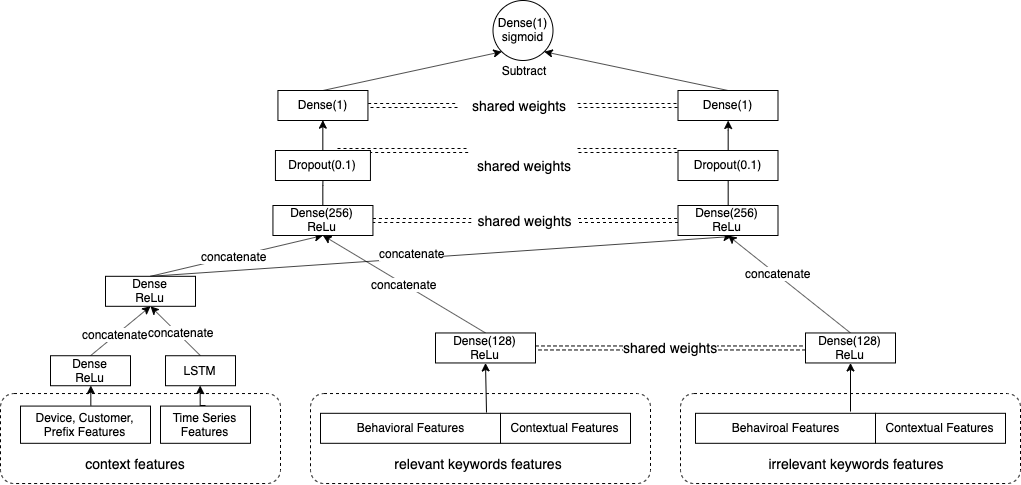}
\caption{Architecture of DeepPLTR}
\label{fig:DeepPLTR Architecture}
\end{figure*}

\subsection{Model Architecture}
Given the extracted pairs, the DeepPLTR architecture is designed as in Figure \ref{fig:DeepPLTR Architecture}. It's a typical RankNet \cite{burges2010ranknet} structure based on fully-connected neural network \cite{sainath2015convolutional}. Different from RankNet \cite{burges2010ranknet}, it adds a context representation layer, which is scalable to host contextual signals. Most of the dense layers are weight-sharing as they represents the same concepts (e.g. query representation, query-context representation). The LSTM \cite{gers1999learning} layer captures the sequential signal of the time series related features, which has better performance than dense structure. The final output layer applies the sigmoid activation \cite{han1995influence} function and binary cross-entropy loss \cite{de2005tutorial} by subtracting the two ranking scores from previous layers. After fine-tuning the parameters through a hyper-parameter tuning process, query representation layer has 128 neurons, LSTM has 16 neurons, context representation layer has 128 neurons, and dropout layer ratio is 0.1. All dense layers apply the ReLU \cite{nair2010rectified} activation function. The output labels are set to 1, as the model is trained based on positive keywords ranking higher than negative keywords.

\subsection{Ranking Loss}
Ranking loss is composed of 3 parts: binary cross entropy loss, $|\Delta NDCG|$ and sample weights. Binary cross entropy loss represents the loss when reversing the ranking order of positive-negative pair. Sample weights represent the optimization target, for example, if the model optimizes GMV, we set GMV as the weight of each pair. $|\Delta NDCG|$ considers the change in NDCG when swapping the ranking positions of positive and negative keywords, so that a keyword pair more separated in the ranking list gets a larger loss. Then we define the loss as below:

$S$ denotes the full set of pairs from Section 3.1. $p$ denotes the positive keywords and $n$ denotes the negative keywords. $w_{p,n}$ denotes the sample weight of the pair. $f$ is the learned ranking function which minimizes the ranking loss. $f_p$ and $f_n$ denote the ranking score of the keywords $p$ and $n$. $|\Delta NDCG_{p,n}|$ denotes the change in NDCG when swapping the positions of keywords $p$ and $n$. $rank_p$ denotes the ranking position of the keywords $p$ ($rank_p \in [1, 2, 3, ...]$. $y$ denotes the output label. $P_{p,n}$ denotes the probability that the positive keyword $p$ is ranked above the negative keyword $n$, which serves the purpose of a contrastive loss that helps distinguish the positive keyword from the negative keyword: when the difference between $f_p$ and $f_n$ is larger, $P$ will be larger. Then we have:

\begin{equation*}
\resizebox{0.45\textwidth}{!}{$L = - \frac{1}{|S|} \sum_{(p,n) \in S} \Biggl[ (y \log P_{p,n} + (1 - y) \log (1-P_{p,n})) \cdot |\Delta NDCG_{p,n}| \cdot w_{p,n} \Biggl]$}
\label{eq:1}
\end{equation*}

where $P_{p,n}$ is defined as
\begin{equation}
    P_{p,n} = \frac{1}{1 + e^{-(f_p - f_n)}}
\end{equation}

and $|\Delta NDCG_{p,n}|$ is defined as:
\begin{equation}
    |\Delta NDCG_{p,n}| = |\frac{2^{rel_p} - 1}{\log (rank_p + 1)} - \frac{2^{rel_p} - 1}{\log (rank_n + 1)}|.
\end{equation}

Given $y$ is always 1 and $rel_p=1$, we simplify the loss from Eq. \eqref{eq:1} as:

\begin{equation}
\label{eq:2}
\resizebox{0.42\textwidth}{!}{$L = - \frac{1}{|S|} \sum_{(p,n) \in S} \Biggl[ (\log \frac{1}{1 + e^{f_n - f_p}}) \cdot |\frac{1}{\log (rank_p + 1)} - \frac{1}{\log (rank_n + 1)}| \cdot w_{p,n} \Biggl]$}
\end{equation}

\subsection{Query Representation And Contextual Features}

\begin{figure*}[b]
\centering
\includegraphics[width=\linewidth]{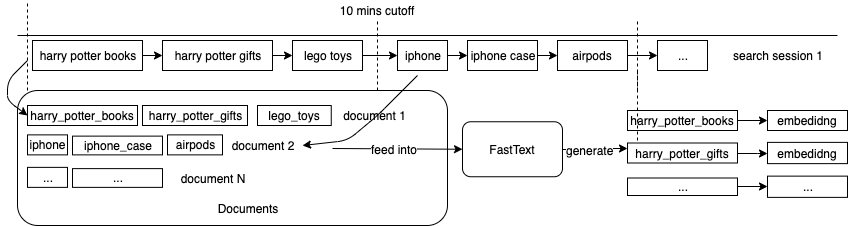}
\caption{Query Embedding Generation Process}
\label{fig:Query Representation}
\end{figure*}

We construct contextual features by calculating the cosine distance between candidate queries and past search queries using pre-trained query embeddings. In this paper, we use the publicly available $fastText$ library \cite{bojanowski2017enriching} to learn the query embeddings. We use the skipGram model where the goal is to independently predict the presence or absence of the context words. The problem is framed as a binary classification task. For the word at position $t$, $fastText$ considers all context words as positive examples and sample random negatives words from dictionary as described in \cite{bojanowski2017enriching, mikolov2013distributed}. It uses the negative log likelihood, $l : x \to log( 1 + e^{-x})$ as the binary logistic loss. The objective function is defined as:

\begin{equation}
\sum_{t=1}^{T} \Biggl[ \sum_{c \in Context} l(s(w_t,w_c)) + \sum_{n \in N_{t,c}} l(-s(w_t,n)) \Biggr]
\end{equation}

where $w_t$ is the target word, $w_c$ is the context word, $C$ is context words, and $N_{t,c}$ is a set of negative examples sampled from the vocabulary. The scoring function $s(w,c)$ is the scalar product between word and context vectors: $s(w_t , w_c) = u_{w_t}^T v_{w_c} $, where vectors $u_{w_t}$ and $v_{w_c}$ corresponding to words $w_t$ and context words $w_c$ respectively.

Our goal is to learn scalable embeddings across different marketplaces and languages. For US, we collect the past 60 days' search sessions (a sequential search actions), and divide each session into multiple sequences of queries, such that no consecutive queries in a sequence are separated by > 10 minutes. We normalize queries by removing special characters and replace space character ‘\ ’ to ‘\_’. We treat each query as a “word”, and sequences of queries as a “document”, and feed them into $fastText$ \cite{bojanowski2017enriching} to train and generate embeddings with 50 dimensions. We call it the {\em whole query embedding} approach. Figure \ref{fig:Query Representation} illustrates the whole query embedding generation process. Compared to the token embedding approach, whole query embedding is a scalable approach as it generates query embeddings WW without language specific processing, e.g. tokenization for Chinese and Japanese. Then at online, we apply the pre-trained embeddings to calculate the cosine similarity between each candidate and past search query as contextual features. In particular, given recently searched query $Q$ and candidate $S$, query embedding $vec()$. $cosine(S, Q) = \frac{vec(S) \cdot vec(Q)}{|vec(S)| \cdot |vec(Q)|}$.

\section{Experiments}
\subsection{Dataset and Experiment Settings}
To conduct experiments, we follow the labeling strategy described in Section 3.1 to extract 77M training pairs, and 11M validation pairs from AC sessions which lead to purchase events (GMV $> 0$) during a week of US data in 2020. This gives us a good coverage of user context features to learn deep model. In order to benchmark on MRR and NDCG, we collect 150k impression logs with ground truth as the test datasets in Feb. 8, 2020. Different models are trained and evaluated with the same sampled data.

\subsection{Benchmark Rankers}
To evaluate DeepPLTR, we constructed 4 benchmark rankers:

\noindent \textbf{MPC Ranker:} the MostPopularCompletion ranker \cite{bar2011context}, which is a common benchmark for AC task. It ranks queries by the decayed search popularity of the query.

\noindent \textbf{MPGC Ranker:} the MostPopularGMVCompletion ranker, which ranks by the decayed GMV of query. It's designed to benchmark with GMV-optimized model.

\noindent \textbf{LambdaMART Ranker (context-aware):} train a LambdaMART ranker using XGBoost library \cite{chen2016xgboost} with the same features as DeepPLTR. From which, we will learn whether NN architecture could improve performance. We set pairwise loss as the optimization target, and fine-tuned number of trees as 150 and max depth as 6 to XGBoost to train the model. We treat this as a baseline ranker.

\noindent \textbf{DeepPLTR-NDCG Ranker (context-aware):} train a DeepPLTR excluding $|\Delta NDCG|$ loss in the Eq. \ref{eq:2} to learn the position impact.

\subsection{Evaluation Metrics}
\noindent \textbf{MeanReciprocalRank:} the average of the reciprocal rank of the target queries in the AC results weighed on the business metrics. Given a test dataset $S$, the MRR for ranker $A$ is computed as:
\begin{equation}
    MRR(A) = \frac{1}{|S|} \sum_{C_t,q_t \in S} \frac{w}{hitRank(A,C_t,q_t)},
\end{equation}

where $C_t$ represents the user context at time $t$, $q_t$ represents the target query, and $w$ represents the sample weight. The function $hitRank$ computes the rank of the relevant query based on the order created by algorithm $A$. The relevant query refers to the clicked query in AC.

\noindent \textbf{NormalizedDiscountedCumulativeGain:} Discounted Cumulative Gain ($DCG$) represents the usefulness or gain of the query based on its position in the ranked list. Given AC task having one relevant query, $IDCG$ will always be 1, so NDCG@p can be simplified as below:

\begin{equation}
    NDCG_p = \frac{\sum_{s \in S} \Biggl[ w_s \cdot (\sum_{i=1}^{p} {\frac{2^{rel_i} - 1}{\log (i + 1)})} \Biggl]}{\sum_{s \in S} w_s},
\end{equation}

where $i$ denotes the rank, $rel_i \in \{0,1\}$ is the relevance of query at rank $i$, and $w_s$ is the sample weight.

\subsection{Results}

To learn the impact of contextual signals on the model's performance, we construct three versions of test samples: All Samples (AS), Samples With Past Searches (SWPS) and Samples Without Past Searches (SWOPS). Evaluation metrics are set to MRR, NDCG@1, NDCG@3.

\begin{table}
  \caption{Evaluate on All Samples}
  \begin{tabular}{l | c | c | c }
     \toprule
     Model & MRR & NDCG@1 & NDCG@3 \\
     \midrule
     MPC & 0.469(-16.84\%) & 0.293(-22.49\%) & 0.464(-18.74\%) \\
     MPGC & 0.488(-13.48\%) & 0.308(-18.52\%) & 0.485(-15.06\%) \\
     LambdaMART & 0.564 & 0.378 & 0.571 \\
     DeepPLTR-NDCG & 0.577(+2.30\%) & 0.400(+5.82\%) & 0.584(+2.28\%) \\
     DeepPLTR & 0.586(\textbf{+3.90\%}) & 0.409(\textbf{+8.20\%}) & 0.594(\textbf{+4.03\%}) \\
     \bottomrule
  \end{tabular}
  \label{tab:Evaluate on All Samples}
\end{table}

\begin{table}
  \caption{Evaluate on Samples WithOut Past Searches}
  \begin{tabular}{l | c | c | c }
     \toprule
     Model & MRR & NDCG@1 & NDCG@3 \\
     \midrule
     MPC & 0.481(-16.20\%) & 0.297(-23.06\%) & 0.474(-18.56\%) \\
     MPGC & 0.505(-12.02\%) & 0.315(-18.39\%) & 0.502(-13.75\%) \\
     LambdaMART & \textbf{0.574} & 0.386 & \textbf{0.582} \\
     DeepPLTR-NDCG & 0.570(-0.70\%) & 0.385(-0.26\%) & 0.580(-0.34\%) \\
     DeepPLTR & 0.572(-0.35\%) & 0.388(\textbf{+0.52\%}) & 0.581(-0.17\%) \\
     \bottomrule
  \end{tabular}
  \label{tab:Evaluate on Samples With No Past Searches}
\end{table}

\begin{table}
  \caption{Evaluate on Samples With Past Searches}
  \begin{tabular}{l | c | c | c }
     \toprule
     Model & MRR & NDCG@1 & NDCG@3 \\
     \midrule
     MPC & 0.459(-17.45\%) & 0.289(-22.10\%) & 0.455(-19.04\%) \\
     MPGC & 0.474(-14.75\%) & 0.303(-18.33\%) & 0.470(-16.37\%) \\
     LambdaMART & 0.556 & 0.371 & 0.562 \\
     DeepPLTR-NDCG & 0.582(+4.68\%) & 0.413(+11.32\%) & 0.587(+4.45\%) \\
     DeepPLTR & 0.598(\textbf{+7.55\%}) & 0.427(\textbf{+15.09\%}) & 0.606(\textbf{+7.83\%}) \\
     \bottomrule
  \end{tabular}
  \label{tab:Evaluate on Samples With Past Searches}
\end{table}

Evaluation on AS are shown in Table \ref{tab:Evaluate on All Samples}. LambdaMART ranker outperforms the naive rankers (MPC, MPGC), which is expected with more advanced signals. DeepPLTR outperforms the LambdaMART by 3.90\% in MRR, by 8.20\% in NDCG@1, and by 5.99\% in NDCG@3, showing that NN-based architecture imrpoves the performance. DeepPLTR outperforms DeepPLTR-NDCG, shows that considering position factor in ranking boosts the ranking performance.

Evaluation on SWOPS are shown in Table \ref{tab:Evaluate on Samples With No Past Searches}. LambdaMART ranker slightly outperforms DeepPLTR on $NDCG@3$, which shows that DeepPLTR has similar performance to LambdaMART ranker without past search history. LambdaMART ranker slightly outperforms PLTR, showing that device type feature improves performance. Evaluation on SWPS are shown in Table \ref{tab:Evaluate on Samples With Past Searches}. DeepPLTR outperforms baseline by 7.55\% in MRR and 15.09\% in NDCG@1, showing that DeepPLTR can learn implicit ranking features based on past searches features. DeepPLTR boosts performance the most on SWPS samples.

We also conduct an online A/B experiment to compare the DeepPLTR and LambdaMART ranker, experiment shows DeepPLTR yields a +0.06\% (p < 0.1) GMV lift compared to LambdaMART ranker, which demonstrates the DeepPLTR outperforms LambdaMART.

\section{Conclusions and Future Work}
In this paper, we have proposed a neural-network based pairwise ranker and ranking loss, and described a scalable approach to learn query embeddings using search session data. The architecture of DeepPLTR is scalable for accommodating a variety of ranking signals and can be applied to other search tasks such as product search. We conduct offline and online experiments to verify that DeepPLTR outperforms LambdaMART ranker. In the future, we will improve AC in several directions: 1) Incooporate richer contextual information, e.g. past product clicks \& purchases; 2) Add exploration mechanism to explore unshown suggestions.

\begin{acks}
We would like to thank the team members in Amazon Search: Trishul Chilimbi, Adam Kiezun, Wenyang Liu, Hsiang-Fu Yu, Erte Pan, Yun Hyokun, Zia Hasan, Pratik Lahiri, RJ, Yukun Lin, Haining Yu, Ben Wing, Hye Jin Jang for their support.
\end{acks}

\bibliographystyle{ACM-Reference-Format}
\bibliography{references}
\end{document}